\documentclass[pdflatex,oneside,sn-mathphys-num]{sn-jnl}
\usepackage{float}
\usepackage{booktabs}
\usepackage{threeparttable}
\usepackage{siunitx}
\usepackage{longtable}
\usepackage{graphicx}
\usepackage{placeins}
\usepackage{multirow}%
\usepackage{amsmath,amssymb,amsfonts}%
\usepackage{amsthm}%
\usepackage{mathrsfs}%
\usepackage[title]{appendix}%
\usepackage{xcolor}%
\usepackage{textcomp}%
\usepackage{manyfoot}%
\usepackage{booktabs}%
\usepackage{algorithm}%
\usepackage{algorithmicx}%
\usepackage{algpseudocode}%
\usepackage{listings}%
\usepackage{tabularx}

\usepackage{array}
\newcolumntype{L}{>{\raggedright\arraybackslash}X}
\theoremstyle{thmstyleone}%
\theoremstyle{thmstyletwo}%
\theoremstyle{thmstylethree}%
\usepackage{caption}

\usepackage{adjustbox}
\usepackage{booktabs}

\begin{document}

\title[Group Differences in Psychopathology Networks]{Rethinking Group Differences in Psychopathology Networks: A Slow–Fast Perspective on Context and Symptom Activation}

\author*[1]{\fnm{Kyuri} \sur{Park}}\email{k.park@uva.nl}
\author[3]{\fnm{Denny} \sur{Borsboom}}\email{d.borsboom@uva.nl}
\author[1,2]{\fnm{Mike} \sur{Lees}}\email{m.h.lees@uva.nl}
\author[4]{\fnm{Leonie} \sur{Elsenburg}}\email{l.k.elsenburg@amsterdamumc.nl}
\author[5]{\fnm{Gaby} \sur{Lunansky}}\email{g.lunansky@amsterdamumc.nl}
\author[4]{\fnm{Karien} \sur{Stronks}}\email{k.stronks@amsterdamumc.nl}
\author[1]{\fnm{Johan} \sur{Bollen}}\email{j.l.t.m.bollen@uva.nl}
\author[4]{\fnm{Mary} \sur{Nicolaou}}\email{m.nicolaou@amsterdamumc.nl}
\author[1,2]{\fnm{Vítor V.} \sur{Vasconcelos}}\email{v.v.vasconcelos@uva.nl}

\affil*[1]{\orgdiv{University of Amsterdam, Informatics Institute},
\orgname{Computational Science Lab},
\orgaddress{\street{PO Box 94323}, \city{Amsterdam},
\postcode{1090GH}, \country{The Netherlands}}}

\affil[2]{\orgdiv{University of Amsterdam},
\orgname{Institute for Advanced Study},
\orgaddress{\street{Oude Turfmarkt 147}, \city{Amsterdam},
\postcode{1012GC}, \country{The Netherlands}}}

\affil[3]{\orgdiv{University of Amsterdam},
\orgname{Department of Psychology},
\orgaddress{\street{Nieuwe Achtergracht 129}, \city{Amsterdam},
\postcode{1001 NK}, \country{The Netherlands}}}

\affil[4]{\orgdiv{Department of Public and Occupational Health, Amsterdam Public Health Research Institute},
\orgname{Amsterdam UMC, University of Amsterdam},
\orgaddress{\street{Van der Boechorststraat 7}, \city{Amsterdam},
\postcode{1081 BT}, \country{The Netherlands}}}

\affil[5]{\orgdiv{Department of Epidemiology and Data Science, Amsterdam Public Health Research Institute},
\orgname{Amsterdam UMC, Vrije Universiteit Amsterdam},
\orgaddress{\street{De Boelelaan 1117}, \city{Amsterdam},
\postcode{1081 HV}, \country{The Netherlands}}}

\abstract{
Groups differing in social and economic circumstances often differ markedly in depressive symptom levels while their estimated symptom-network structures show few clear differences. Null results in network comparisons are interpreted as if indicating that the two groups' symptom systems are the same. However, it only indicates that no difference is detected in the interactions between symptoms. How readily symptoms become present, and how that may depend on the contextual conditions under which symptoms are observed, is rarely compared. Building on a slow--fast perspective, we conceptualize depressive symptoms as a relatively fast-changing system embedded in more persistent social, economic, psychosocial, health, and lifestyle conditions that shape symptom activation.

Using a cohort of 23,689 adults from Amsterdam, we illustrate this perspective with depressive symptoms measured by the PHQ-9 and a composite index of slow-layer contextual conditions, which we call {\em Slow Risk Load} (SRL), capturing persistent socioeconomic, psychosocial, and health-related conditions. Participants with High SRL showed substantially higher depressive symptom levels than those with Low SRL. A standard Network Comparison Test did not detect an overall difference in network structure between the groups, although global strength was modestly higher in the High SRL group.

Ising models, which separate how symptoms co-occur from how readily each becomes present, indicate that activation is higher under high SRL for eight of nine symptoms, and model comparison consistently favored group-specific activation over group-specific interactions.

The slow--fast perspective reframes group network comparisons as a problem spanning multiple time scales in which symptom-network structure, symptom activation, and contextual conditions should be considered jointly. It places symptom-focused and context-focused interventions at different layers of the same coupled system rather than in competition.
}

\maketitle
\section{Introduction}\label{sec1}

Depression is often summarized by a symptom score that is measured through surveys, but symptoms are
not experienced as isolated boxes to be ticked
\cite{fried2015consistent,fried2015sum}. A night of poor sleep can
leave a person exhausted, exhaustion can make daily tasks feel harder,
reduced activity can deepen low mood, and low mood can make sleep more
difficult again. Symptoms are, therefore, not only passive signs of an
underlying disorder. They can also form a self-sustaining system. This
idea lies at the center of network theory of psychopathology, in which
mental disorders are understood as systems of interacting symptoms
rather than as collections of interchangeable indicators of a single
latent disease process
\cite{cramer2010comorbidity,borsboom2017network,cramer2016major,
robinaugh2020network,bringmann2022psychopathological,park2026feedback}.

Building on this network perspective, a growing literature compares
estimated symptom networks across populations. Symptom network models consist of \textit{nodes}, representing the symptoms, and \textit{edges}, which are the connections representing the estimated associations between the symptoms. Network models have been used to compare groups defined by gender \cite{lee2022sex,izquierdo2023sex,alcalde2024how}, age \cite{tao2023comparing}, migration background \cite{elsenburg2026precariousness}, clinical status \cite{vanborkulo2015association}, treatment response \cite{mcelroy2019structure}, and other characteristics. These studies
commonly compare overall connectivity (i.e., the strength of all the edges in the network), specific symptom associations, and node centrality (i.e., which symptom has the most and strongest connections to other symptoms \cite{epskamp2018estimating}).  Their findings can present an interpretive puzzle. Groups may differ clearly in symptom prevalence or severity while showing few or inconsistent differences in estimated network structure \cite{steen2021symptom}. In other cases, differences emerge in global strength, particular edges, or centrality measures\cite{vanborkulo2015association,mcelroy2019structure,lee2022sex}, but their source and substantive meaning remain uncertain. The pattern in which connectivity remains largely invariant while symptom prevalence or severity differs between groups has also been reported across different populations and study designs \cite{steen2021symptom,elovainio2021symptom,vandertuin2023relating}.

The fundamental challenge is that an estimated network comparison is
not self-explanatory. Some apparent differences may reflect sampling
error, measurement noise, model specification, or other forms of
heterogeneity
\cite{epskamp2018estimating,epskamp2017estimating,borsboom2021network,
briganti2024network}. Even when a difference is stable under reasonable
robustness checks, an estimated statistical association does not by
itself establish that one symptom causally affects another or that
causal symptom relations differ across groups
\cite{ryan2022challenge,park2024discovering,park2026mechanistic}. Network comparisons require an additional interpretive step between the observed
statistical pattern and the symptom-generating processes proposed to
underlie it.

For binary symptom data, the Ising model is particularly useful for making this distinction because it describes the observed distribution using two sets of parameters: pairwise interaction
parameters (i.e., the \textit{edge parameters}) which represent conditional statistical associations among
symptoms, and activation or external-field parameters (i.e., the \textit{node thresholds}), which represent
the conditional tendency for individual symptoms to be present
\cite{van2014new,kruis2016three,marsman2018introduction,
haslbeck2021interpreting}.

\begin{figure}[!htbp]
\centering
\includegraphics[width=1\textwidth]
{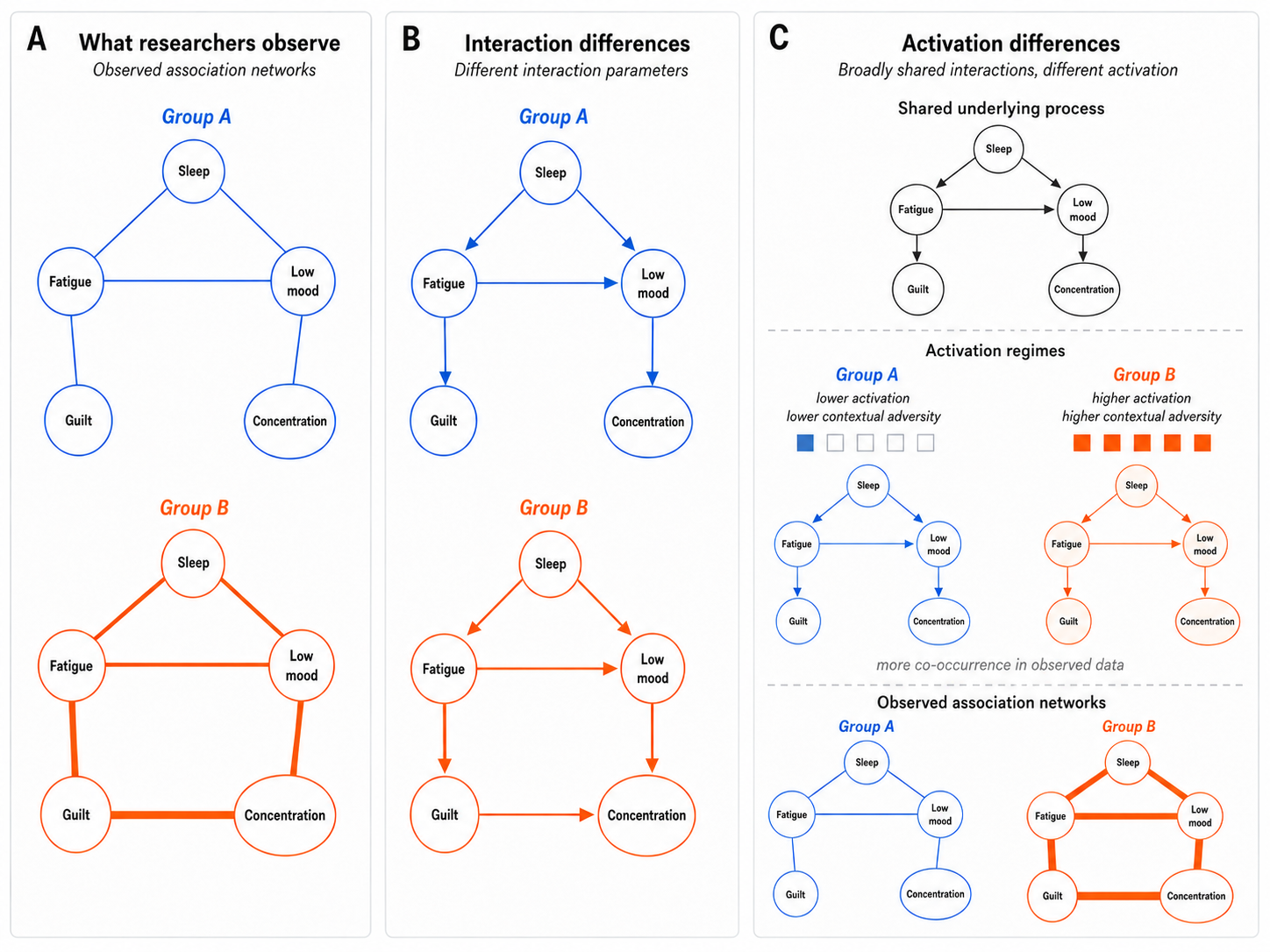}
\caption{\textbf{Interaction and activation components of group
differences in estimated symptom networks.}
\textbf{(a)} Estimated association networks can differ across groups,
for example by appearing weaker or stronger. These networks summarize
statistical associations in observed symptom data and do not by
themselves identify causal symptom effects.
\textbf{(b)} One limiting case is that the processes represented by
the symptom-interaction parameters differ across groups. In the
schematic example, Group B has an additional process feature that may
induce an additional observed association between guilt and
concentration. The arrows are a schematic of an inferred causal model.
\textbf{(c)} A second limiting case is that the interaction structure
is broadly shared, while groups differ in symptom activation because
they are observed under different contextual conditions. Higher
activation changes symptom prevalence and co-occurrence, although it
does not necessarily imply different interaction parameters in a
correctly specified Ising model in which activation is freely
estimated. Empirical group differences may involve either component
or both components simultaneously.}
\label{fig:two_explanations}
\end{figure}

Figure~\ref{fig:two_explanations} illustrates how group differences may appear in these two components. In one limiting case, groups differ in the processes represented by the interaction parameters (the connections or edges). In another,
the interaction structure is broadly similar across groups, but symptoms are more easily activated in one group than in the other (i.e., the nodes have different \textit{threshold parameters} to become active). These possibilities are not mutually exclusive: groups may differ in interaction parameters, activation parameters, or both.\footnote{Distinguishing between these possibilities is not simply a matter of estimating the two components separately. In the $\{0,1\}$ Ising parameterization, interaction and activation estimates are statistically interdependent, so constraining one component to equality across groups can affect the estimated differences in the other \cite{finnemann2026theory}. We therefore test equality of both interaction and activation parameters rather than assuming either to be shared across groups.}

Activation parameters can differ for many reasons, including
biological, psychological, and environmental
factors. Slowly varying contextual conditions are one theoretically
important source. Social and economic resources, work and housing
conditions, chronic stress exposure, physical health, and lifestyle
constraints have long been recognized as important determinants of
depressive symptoms and other mental health outcomes
\cite{pearlin1981stress,lund2018social,alegria2018social,
compton2015social,patel2018lancet,ridley2020poverty,
kirkbride2024social,elsenburg2026precariousness,park2026precariousness}. Much of that work examines particular contextual
factors and their associations with mental health outcomes. What is
harder is to represent symptoms and contextual conditions together as
parts of one coupled system. One reason is that they often operate on
different time scales: symptoms such as depressed mood, fatigue, sleep
problems, and concentration difficulties can fluctuate over hours,
days, or weeks, whereas many contextual conditions persist over months
or years.

\begin{figure}[htbp]
\centering
\includegraphics[width=0.80\textwidth]{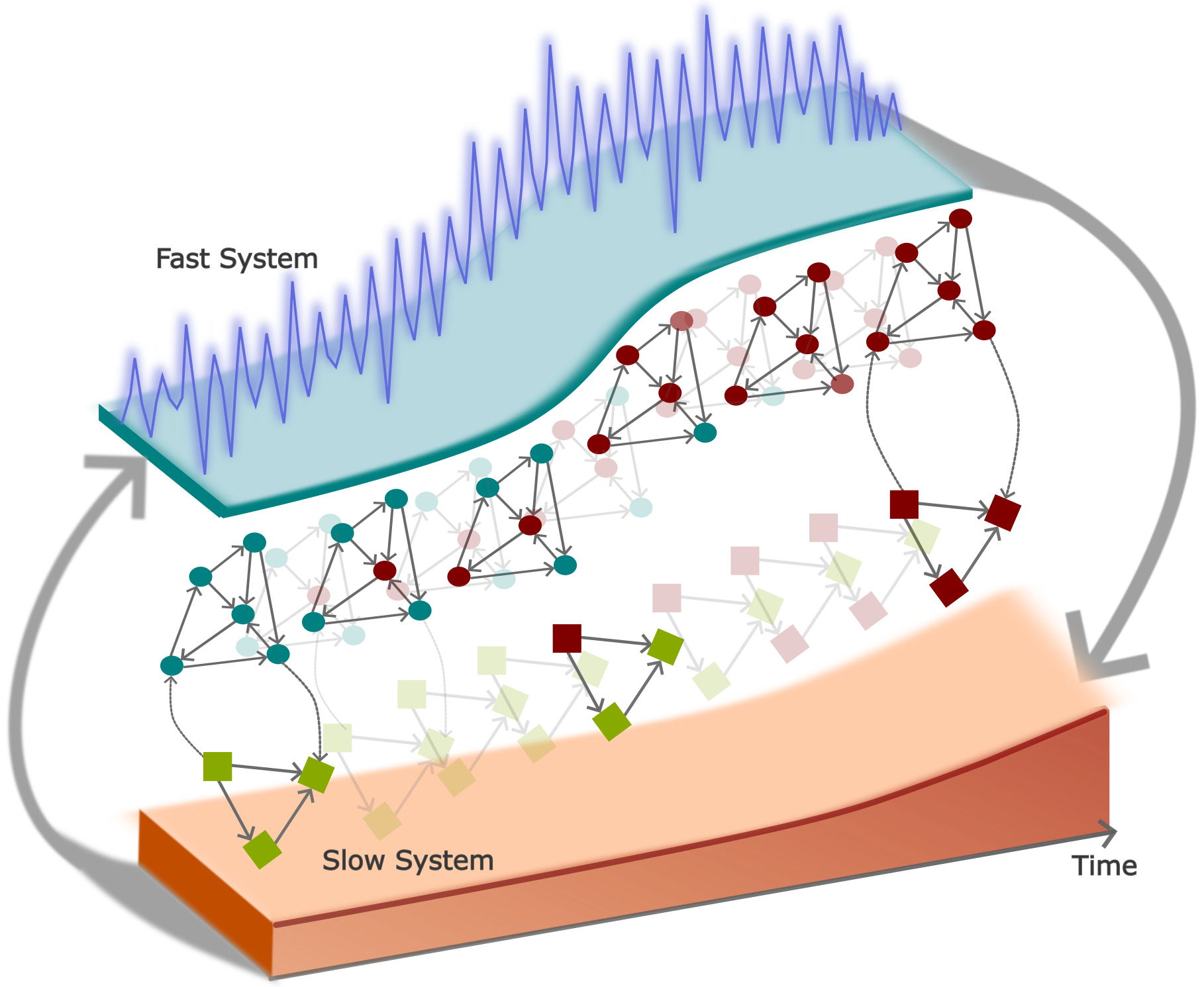}
\caption{\textbf{Coupled slow--fast perspective on symptom dynamics.}
Symptoms form a relatively fast-changing system, while contextual
conditions often evolve more slowly. The slow layer includes social,
economic, psychosocial, health-related, and lifestyle conditions that
may shift how easily symptoms become active. Moving the fast system
toward higher activation makes symptoms more likely to be present and
to co-occur. The relationship is bidirectional: long-term, sustained symptom
activation may also feed back into the slow layer over time.}
\label{fig:slowfast_concept}
\end{figure}

Building directly on the slow--fast framework developed by Lunansky
and colleagues, this multiscale picture can be organized in terms of
a relatively fast symptom layer and a more slowly changing contextual
layer \cite{lunansky2020personality}. The fast layer represents
symptoms and their mutual relations. The slow layer represents more
persistent contextual conditions that may shift how easily symptoms
become active. This framing does not deny the importance of individual
contextual factors. Rather, it asks how multiple slower-moving
conditions may jointly position a symptom system closer to, or farther
from, a high-activation state, consistent with dynamical-systems
accounts of depression and resilience
\cite{cramer2016major,van2014critical}.
Figure~\ref{fig:slowfast_concept} illustrates this coupled system
schematically.

The relationship between the two layers is not necessarily
one-directional. So far, we have emphasized how slower-moving
contextual conditions may shape symptom activation. Sustained symptom
activation may also feed back into the slow layer over time.
Long-lasting depressive symptoms can reduce social participation, work
capacity, physical health, financial stability, or access to
supportive environments
\cite{lund2010poverty,ridley2020poverty,park2026precariousness}. These fast-to-slow
connections require a persistent state of the fast system and worsen adverse contextual conditions,
while slow-to-fast effects keep symptoms easier to activate. This
bidirectional structure is central to the theoretical interpretation
of the slow--fast perspective, although the cross-sectional analysis
presented here assumes but cannot establish the direction of influence.

We illustrate this perspective using depressive symptoms and contextual conditions measured in the HELIUS cohort \cite{snijder2017cohort}. To summarize the empirical slow layer, we construct a composite index of relatively persistent social, economic, psychosocial, health-related, and lifestyle conditions, which we call \textit{Slow Risk Load}, SRL. Higher values indicate more adverse slow-layer conditions. We focus on comparing participants with Low and High SRL, while using ethnicity and age as secondary descriptive comparisons.

The empirical analysis focuses on one part of the broader bidirectional system: whether differences in slow contextual conditions are
reflected in symptom-interaction parameters, symptom-activation parameters, or both. More broadly, the aim is not to replace network explanations with contextual ones, but to show how symptom-network structure, symptom activation, and slowly varying contextual conditions can be considered jointly when interpreting group differences.

\section{Results}\label{sec2}

The analysis included 23,689 participants from the HELIUS cohort, a large population-based study in Amsterdam \cite{snijder2017cohort}. Depressive symptoms were measured with the
nine-item Patient Health Questionnaire (PHQ-9)
\cite{kroenke2001phq}.
The central empirical question is whether differences in depressive symptom expression across levels of Slow Risk Load are reflected in symptom--symptom interactions, symptom-activation parameters, or both.

We first describe depressive symptom levels across Slow Risk Load, ethnicity, and age groups. We then focus on the Low--High SRL comparison, beginning with a standard permutation-based comparison of network structure and global strength. Next, we use exact multigroup Ising models to test equality of the interaction and activation
parameters separately. We then examine activation differences when interactions are estimated separately in each group. Finally, we treat SRL as a continuous predictor and assess whether its association differs across symptoms.

\subsection{Slow Risk Load and depressive symptom levels align most clearly in the Slow Risk comparison}

\begin{figure}[htbp]
\centering
\includegraphics[width=0.8\textwidth]{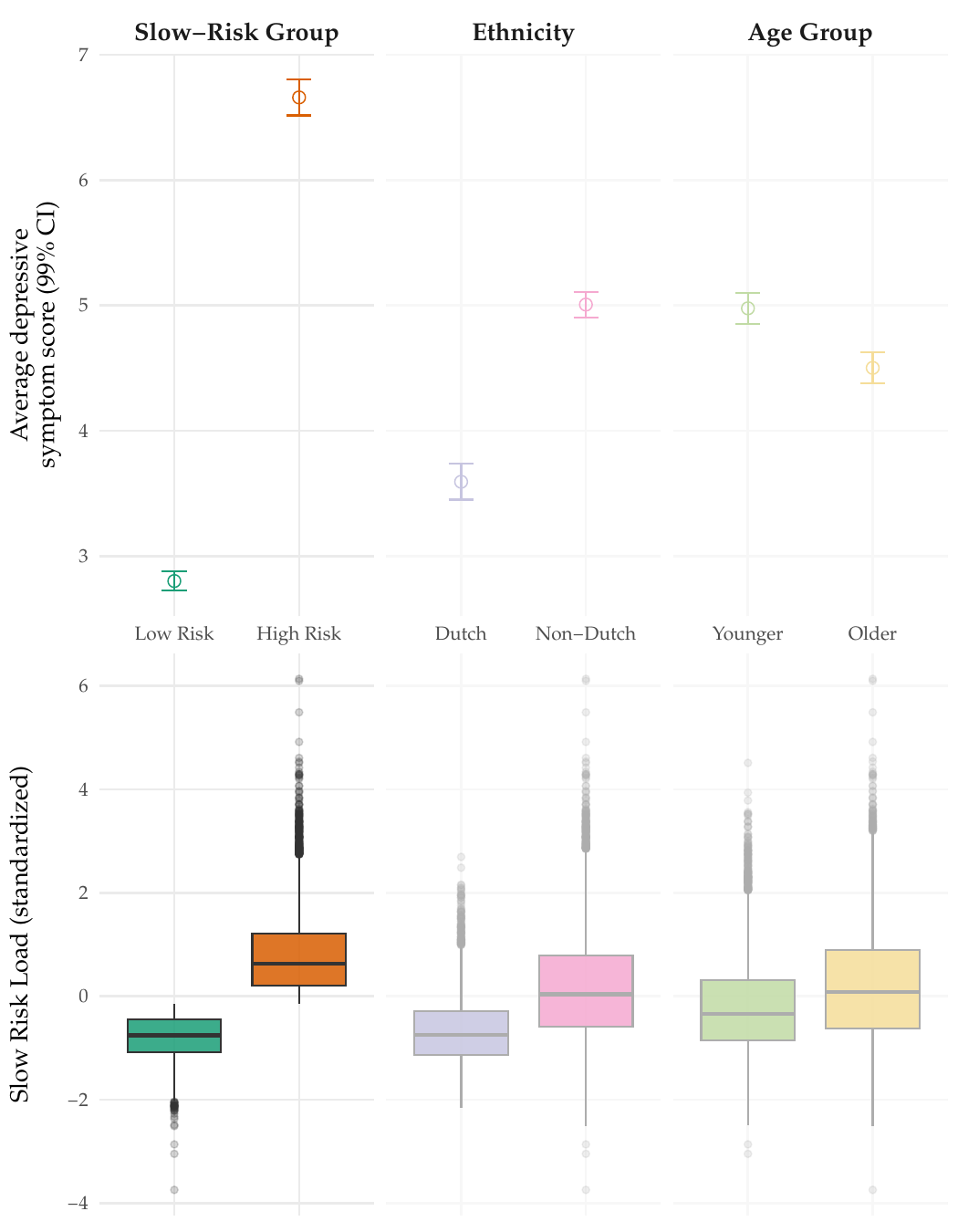}
\caption{\textbf{Depressive symptom level and Slow Risk Load across group comparisons.}
\textbf{Top row}: average depressive symptom score measured with the nine-item Patient Health Questionnaire (PHQ-9). Points show group means and error bars show 99\% confidence intervals.
\textbf{Bottom row}: distributions of \textit{Slow Risk Load}, a standardized composite of relatively persistent contextual conditions; higher values indicate more adverse slow-layer conditions. Boxes show medians and interquartile ranges (IQRs), with whiskers extending to $1.5\times$IQR; points indicate individual observations.
The Slow Risk comparison is the main empirical comparison and is shown in full color, whereas ethnicity and age group are included as secondary descriptive comparisons and are shown in lighter tones. The High Slow Risk group shows both higher depressive symptom level and higher Slow Risk Load. Ethnicity shows a similar but less distinct pattern, whereas the age-group comparison shows little separation in Slow Risk Load and a different pattern for depressive symptom level.}
\label{fig:slow_burden_phq}
\end{figure}

Figure~\ref{fig:slow_burden_phq} summarizes depressive symptom level and Slow Risk Load across the three group comparisons. The Slow Risk
comparison shows the clearest separation. By construction, the High SRL group is shifted upward relative to the Low Slow Risk group
and shows a longer upper tail in Slow Risk Load
(Fig.~\ref{fig:slow_burden_phq}, bottom-left). The same ordering is visible in depressive symptom level: average depressive symptom scores
are higher in the High SRL group than in the Low SRL group (Fig.~\ref{fig:slow_burden_phq}, top-left).

Ethnicity and age group are included as secondary descriptive comparisons. Ethnicity shows a similar, though less distinct, pattern: Non-Dutch participants tend to have higher SRL values than Dutch participants, and average depressive symptom scores are also higher in the Non-Dutch group (Fig.~\ref{fig:slow_burden_phq}, middle column). In contrast, the Younger--Older comparison shows little separation in SRL and a different pattern for depressive symptom level: if anything, the Older group has slightly lower average depressive symptom scores despite somewhat higher SRL values (Fig.~\ref{fig:slow_burden_phq}, right column).

These descriptive comparisons show the clearest alignment for the SRL grouping: the groups most clearly separated on the slow layer also
differ most clearly in depressive symptom level. This pattern does not, however, establish directionality or reveal how that difference is
represented in the fitted symptom system. We therefore ask next whether Low--High SRL groups
differ in overall network structure and global strength.

\subsection{Standard network comparison suggests broadly similar
network structure}

We first compare the Low and High SRL groups using the Network Comparison Test (NCT), a permutation-based procedure for comparing psychological networks across groups \cite{van2023comparing}. Despite the substantial difference in depressive symptom levels, the NCT does not reject overall network-structure invariance between the two groups ($M = 0.319$, $p = .378$). However, estimated global strength is modestly higher in the High SRL group than in the Low SRL group ($23.73$ versus $22.62$; $S = 1.116$, $p = .0037$).

Taken together, these results suggest that the two groups did not differ through a clear, broad reorganization of the estimated symptom network. Although estimated global strength was higher in the High SRL group, this difference should not by itself be interpreted as evidence of stronger underlying symptom interactions. The NCT also does not establish that the interaction parameters are identical across groups. This leaves a more specific question open: whether differences between the Low and High SRL groups are reflected in symptom-interaction parameters, symptom-activation parameters, or both.


\begin{figure}[htbp]
\centering
\includegraphics[width=0.92\textwidth]{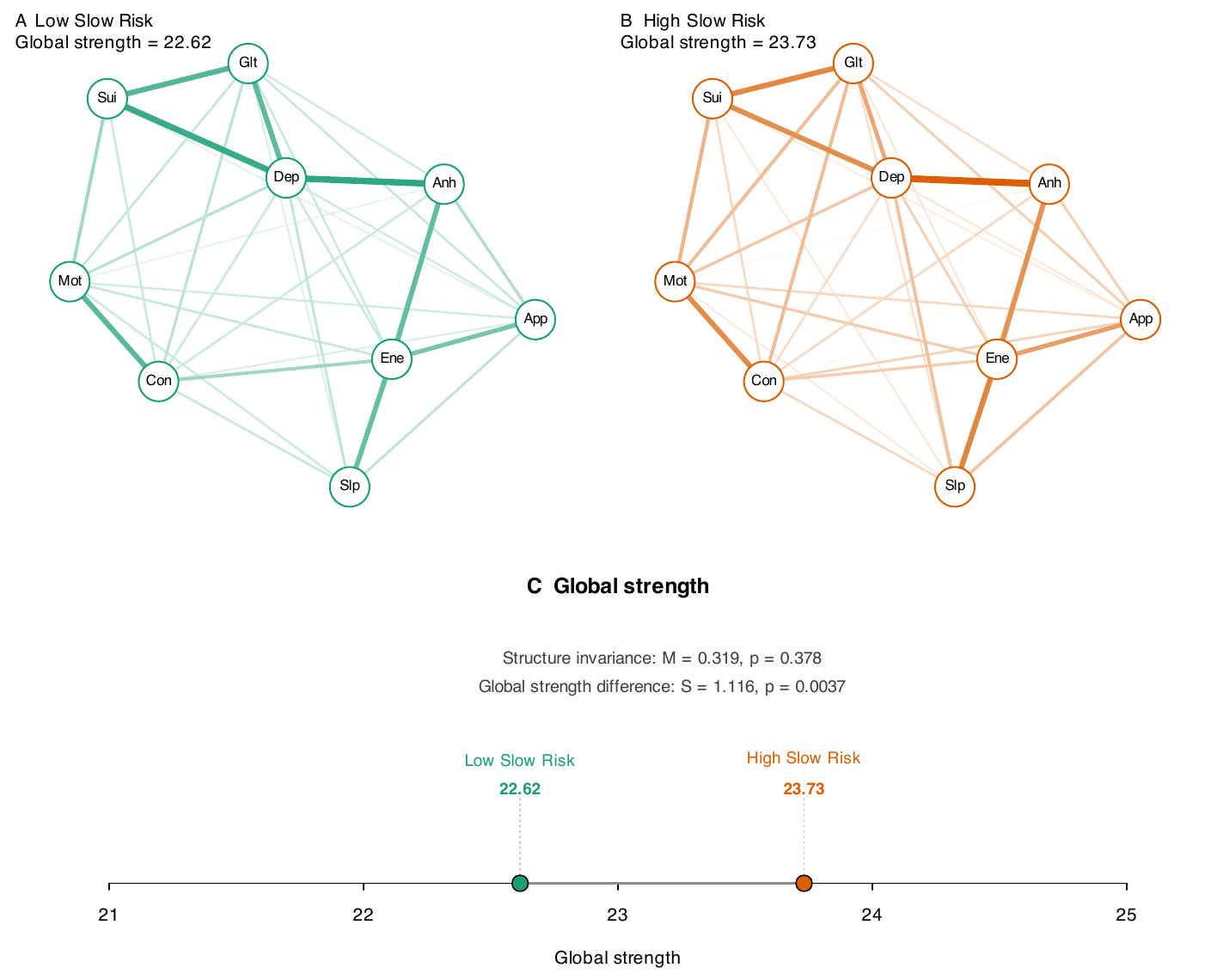}
\caption{\textbf{Standard network comparison for Low and High Slow Risk groups.}
\textbf{(A)} Estimated symptom network for the Low Slow Risk group.
\textbf{(B)} Estimated symptom network for the High Slow Risk group. The two networks are displayed using the same node layout and edge
scale. \textbf{(C)} Global strength in the two groups. The Network Comparison Test did not reject overall network-structure invariance
($M=0.319$, $p=.378$), although global strength was modestly higher in the High Slow Risk group ($S=1.116$, $p=.0037$).}
\label{fig:nct_comparison}
\end{figure}

\subsection{Group differences involve both symptom interactions and
activation}


To examine the difference between High and Low SRL groups more directly, we fit four multigroup Ising models.
The models allow (i) both interaction and activation parameters to differ
between groups, (ii) only activation to differ, and (iii) only
interactions to differ, or (iv) require both sets of parameters to be the
same.

Likelihood-ratio tests rejected exact equality of the interaction parameters across groups, even when activation parameters were allowed to differ, $\chi^2(36)=119.17$, $p<.001$. Likewise, exact equality of the activation parameters was rejected when interactions were allowed to differ, $\chi^2(9)=112.62$, $p<.001$. Given the large sample, however, these tests are highly sensitive to small departures from exact equality and should not be interpreted as evidence that the between-group differences are necessarily large. We therefore consider these results together with the information criteria and the magnitude of the estimated differences.

The information criteria give a more nuanced picture (Table~\ref{tab:ising_model_comparison}). AIC favors the model in which
both interactions and activation differ between groups, whereas BIC
favors the more parsimonious model with common interactions and
group-specific activation parameters. Both criteria also prefer
common interactions with group-specific activation over
group-specific interactions with common activation. The group
difference therefore cannot be attributed entirely to either
component, although, among the restricted models, allowing activation
to differ provided the better account.

The NCT and the multigroup models approach the group comparison from different angles. The NCT looks for a pronounced difference in any single edge between the two regularized networks, whereas the multigroup models ask whether the interaction parameters can be treated as equal when considered together. Viewed side by side, the results
suggest that the overall network structure is broadly similar across
groups, while differences in symptom activation are more apparent.

\begin{table}[htbp]
\centering

\caption{\textbf{Comparison of multigroup Ising models for the Low and
High Slow Risk groups.}
Models differ in whether symptom-interaction and symptom-activation
parameters are estimated separately across groups or constrained to be
equal.}
\label{tab:ising_model_comparison}

\small
\setlength{\tabcolsep}{6pt}
\renewcommand{\arraystretch}{1.12}

\begin{tabular}{@{}lccrrrr@{}}
\toprule
Model
& \(J_{ij}\)
& \(h_i\)
& \(k\)
& LL
& AIC
& BIC \\
\midrule

Fully group-specific
& Free
& Free
& 90
& $-93{,}555.20$
& $\mathbf{187{,}290.40}$
& $188{,}015.39$ \\

Common interactions
& Equal
& Free
& 54
& $-93{,}614.78$
& $187{,}337.57$
& $\mathbf{187{,}772.56}$ \\

Common activation
& Free
& Equal
& 81
& $-93{,}611.51$
& $187{,}385.02$
& $188{,}037.51$ \\

Fully common
& Equal
& Equal
& 45
& $-95{,}229.03$
& $190{,}548.06$
& $190{,}910.55$ \\

\bottomrule
\end{tabular}

\vspace{0.4em}

\begin{minipage}{0.96\textwidth}
\footnotesize
\textit{Note.} \(J_{ij}\) denotes pairwise symptom-interaction parameters,
\(h_i\) denotes symptom-activation parameters, and \(k\) is the number of
freely estimated parameters. ``Free'' indicates that parameters were
estimated separately for the Low and High SRL groups, whereas
``Equal'' indicates that they are constrained to be the same across
groups. LL is the log-likelihood. Higher LL values indicate better fit
before accounting for model complexity. Lower AIC and BIC values
indicate a better balance between fit and complexity. Bold values
indicate the lowest AIC and BIC, respectively.
\end{minipage}

\end{table}

\subsection{Activation differences remain when symptom interactions
are estimated separately}

Because interaction and activation parameters are statistically
interdependent,\footnote{Interaction and activation estimates are not orthogonal in the
$\{0,1\}$ Ising parameterization, constraining the interaction
parameters can affect the estimated activation differences
\cite{finnemann2026theory}.} we now examine activation estimates from the fully
group-specific model, in which both components are estimated
separately for the Low and High SRL groups. We summarize the
group difference for symptom $i$ as
\[
\Delta h_i =
h_{i,\mathrm{High}} - h_{i,\mathrm{Low}},
\]
so that positive values indicate that the symptom has a higher baseline tendency to be present in the High SRL group, conditional on the remaining symptoms being absent.

\begin{figure}[htbp]
\centering
\includegraphics[width=1\textwidth]
{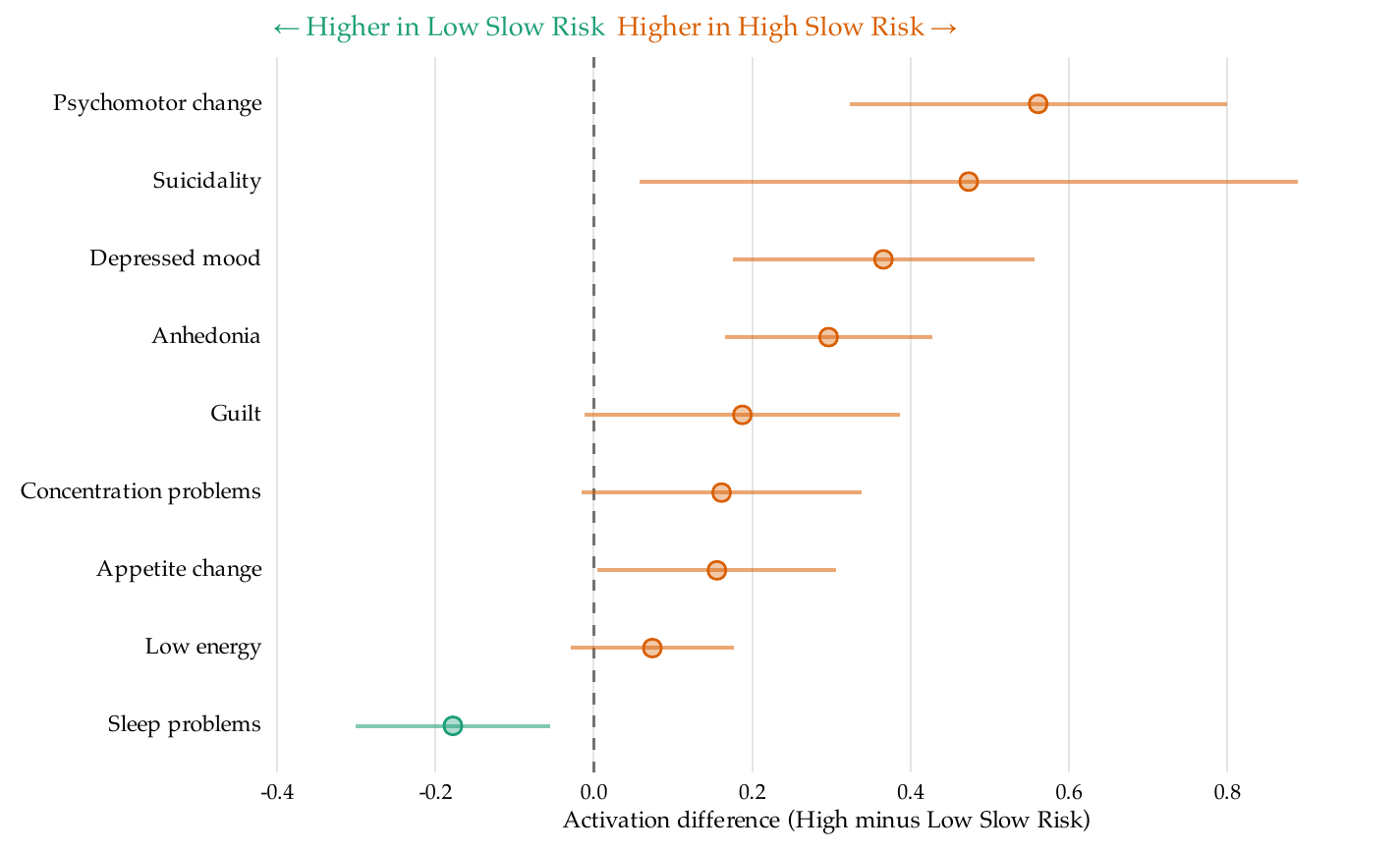}
\caption{\textbf{Differences in symptom-activation parameters between
High and Low Slow Risk Load groups.}
Points show differences in external-field parameters from the fully
group-specific exact Ising model,
$\Delta h_i=h_{i,\mathrm{High}}-h_{i,\mathrm{Low}}$.
Positive values indicate higher activation in the High SRL group, whereas negative values indicate higher activation in the Low
SRL group. Horizontal bars show unadjusted 95\% Wald confidence intervals; Holm-adjusted
$p$-values for the nine symptom-specific contrasts are reported in
Appendix Table~A2.
Both symptom-interaction and activation parameters are estimated
separately in the two groups. These parameters describe the fitted
binary symptom distribution and should not be interpreted as observed
symptom prevalences or causal effects of Slow Risk Load.}
\label{fig:activation_differences}
\end{figure}

Point estimates are higher in the High SRL group for eight of
the nine symptoms (Fig.~\ref{fig:activation_differences}). The largest
positive differences are observed for psychomotor change
($\Delta h=0.561$, 95\% CI $[0.323,0.799]$), suicidality
($\Delta h=0.473$, 95\% CI $[0.058,0.889]$), depressed mood
($\Delta h=0.366$, 95\% CI $[0.175,0.556]$), and anhedonia
($\Delta h=0.296$, 95\% CI $[0.166,0.427]$). Appetite change also
shows a small positive difference
($\Delta h=0.155$, 95\% CI $[0.005,0.306]$). The confidence intervals
for guilt, concentration problems, and low energy include zero.
Sleep problems differ in the opposite direction: the activation
parameter is lower in the High SRL group
($\Delta h=-0.178$, 95\% CI $[-0.301,-0.055]$). 

After Holm correction across the nine symptom-specific contrasts, differences
remained statistically detectable for psychomotor change, anhedonia, depressed
mood, and sleep problems. The contrasts for suicidality and appetite change did
not remain statistically significant after correction.

As a sensitivity analysis, we repeat the comparison using the more parsimonious model in which interactions are constrained to be the same across groups. Under this specification, activation is higher in the High SRL group for all nine symptoms (Appendix Fig.~\ref{fig:appendix_activation_sensitivity}). Across both model specifications, the overall pattern is the same: symptoms generally have a higher conditional baseline tendency to be present in the High SRL group. The exact symptom-specific differences nevertheless vary across specifications, especially for sleep problems.

\subsection{Associations with Slow Risk Load are symptom-specific}

The preceding analyses focused on differences between the Low and High
SRL groups. We next treat SRL as a continuous
variable to examine whether its association with symptom presence varied
across the nine depressive symptoms.

For each symptom, we fit two logistic regression models. The first
estimates the association between SRL and symptom presence
while adjusting for age, gender, and ethnicity. The second additionally
adjusts for the other eight depressive symptoms. The latter
coefficient therefore represents the conditional association between
SRL and a given symptom after accounting for the remaining
symptom profile.

\begin{figure}[htbp]
\centering
\includegraphics[width=1\textwidth]
{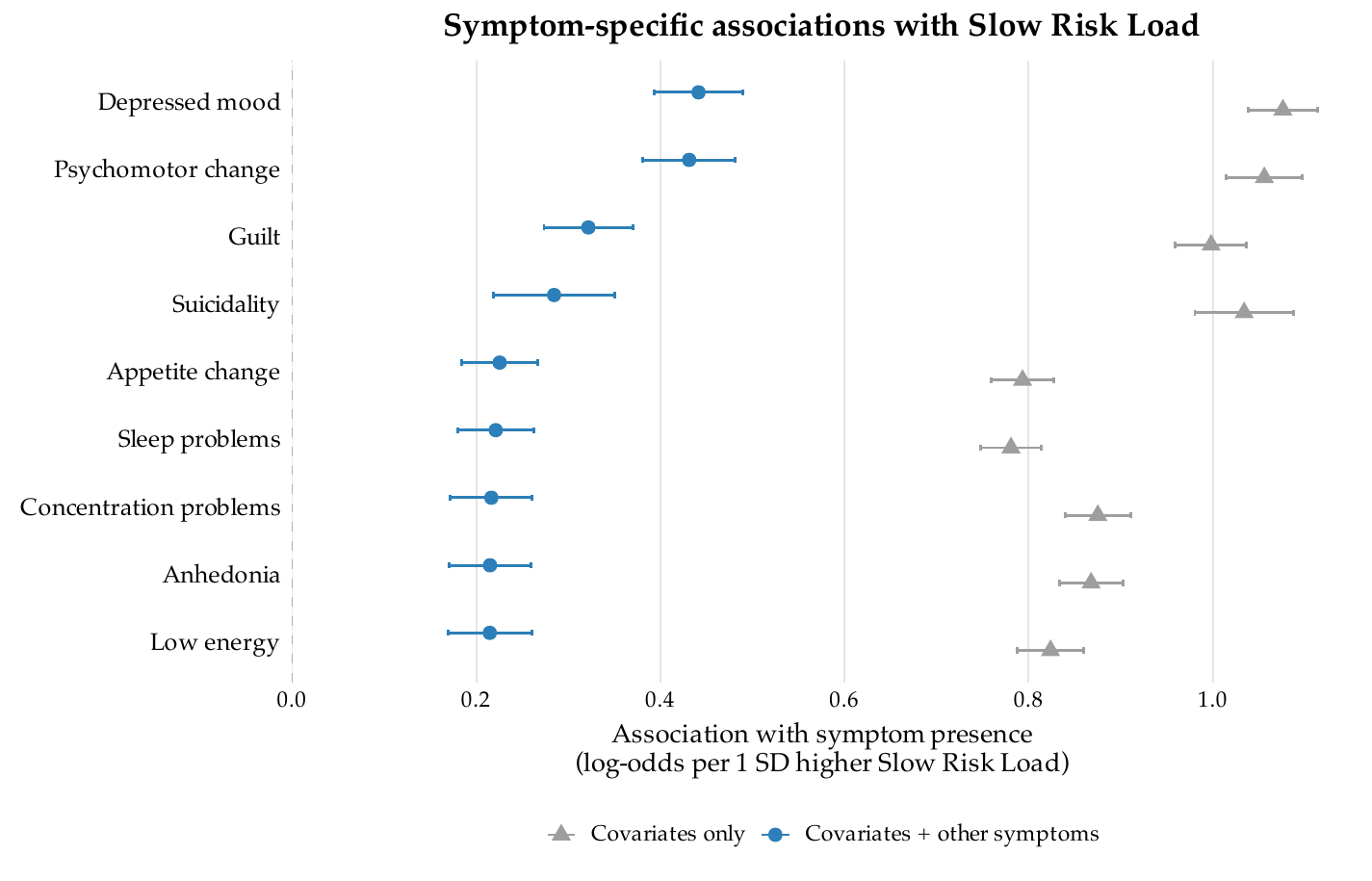}
\caption{\textbf{Associations between continuous SRL and
depressive symptom presence.}
Points show logistic-regression coefficients for the association
between a one-standard-deviation increase in Slow Risk Load and the
presence of each depressive symptom. Covariate-adjusted models adjust
for age, gender, and ethnicity. Symptom-adjusted models additionally
adjust for the other eight depressive symptoms. Horizontal bars show
95\% confidence intervals. The coefficients are conditional
associations and should not be interpreted as causal effects or as
evidence about the direction of relations among symptoms.}
\label{fig:slowrisk_coefficients}
\end{figure}

SRL is positively associated with all nine symptoms in the
covariate-adjusted models
(Fig.~\ref{fig:slowrisk_coefficients}). The estimates are attenuated after adjustment for the other symptoms. Because the other symptoms may themselves lie on pathways linking SRL to a given symptom, these adjusted coefficients should be interpreted descriptively rather than as direct effects of SRL. Nevertheless, the magnitudes of the conditional associations differ across symptoms. Depressed mood and psychomotor change retain the strongest associations, followed by guilt and suicidality. The conditional associations are weaker for
appetite change, sleep problems, concentration problems, anhedonia, and low energy.
These estimates describe symptom-specific conditional associations; they do not establish temporal ordering or causal direction, nor can they distinguish direct associations with contextual conditions from associations arising through other symptoms or unmeasured factors.

\section{Discussion}

The Low and High Slow Risk Load groups differ clearly in depressive symptom expression, but this difference is not accompanied by a clear overall change in symptom-network structure. Instead, our findings point to broadly similar network organization, some differences in symptom interactions, and a more consistent difference in the baseline tendency for symptoms to be present under High Slow Risk Load.

A similar dissociation between connectivity and symptom levels has been reported when comparing groups that differ in depression status or severity directly. Using longitudinal Finnish population data, Elovainio et al. \cite{elovainio2021symptom} found no difference in
network connectivity between individuals with and without diagnosed depression, but markedly lower symptom threshold parameters (i.e., activation parameters in our framework) among those with depression, and concluded that thresholds, rather than connectivity, may be the more important feature distinguishing the two groups. The authors concluded that symptom network models may need to be expanded with external factors that shape symptom thresholds. The slow--fast perspective offers one way of doing exactly that.

In this sense, the slow--fast perspective complements symptom-focused network models by considering the contextual conditions under which symptom systems are observed. Symptom-focused models describe conditional relations among symptoms, whereas slowly varying contextual measures characterize the broader conditions in which those symptom relations occur. Considering both together can therefore provide a fuller account of how groups with markedly different symptom levels may still show broadly similar network organization. This framing builds directly on the slow--fast perspective developed by Lunansky and colleagues \cite{lunansky2020personality}. More broadly, the framework connects the network perspective on symptom interactions with the public-health perspective on the contextual determinants of mental health.

This perspective builds on well-established findings that social,
economic, psychosocial, health, and lifestyle conditions are associated with depression \cite{pearlin1981stress,lund2018social,alegria2018social,
ridley2020poverty,kirkbride2024social}. It clarifies why those conditions matter when interpreting symptom networks. Slowly varying conditions need not be treated only as background characteristics. They may be substantively relevant to how readily symptoms are expressed within the faster symptom system.

This perspective also broadens the questions that guide network
comparisons. Alongside asking, ``How do symptom relations differ
between groups?'', researchers can ask, ``Are these groups living under different contextual conditions that make symptoms more or less likely to be present?'' The first question concerns the organization of the symptom system. The second concerns the conditions under which that system is operating. Our findings suggest that both components are involved in the differences in depressive symptom expression between the Low and High SRL groups, although differences in activation were more consistent than any broad difference in network organization.

Our symptom-specific analyses add a further point. Slow Risk Load is
associated with all nine symptoms, but the strength of these
associations varies across the symptom profile. Contextual conditions
may therefore be reflected across the symptom system as a whole while
still having stronger descriptive links with some symptoms than with
others. These patterns do not identify where contextual influence
begins or how it may spread through the network, but they provide useful
hypotheses for future longitudinal and experimental work.

The slow--fast perspective is also explicitly bidirectional. Contextual
conditions may shape symptom vulnerability, but sustained symptoms can
in turn affect the conditions in which people live. Persistent
depressive symptoms may reduce work capacity, social participation,
physical health, financial stability, or access to supportive
environments
\cite{ridley2020poverty,lund2010poverty}. Feedback processes more generally can change symptom persistence and the response of a dynamical symptom system \cite{park2026feedback}. Slow-to-fast and fast-to-slow
processes may therefore reinforce one another over time. Although the
present empirical analysis, for its cross-sectional nature, focuses mainly on how contextual conditions, which we assume to be more slowly varying at the individual level,
are reflected in the symptom system, the broader framework treats both
layers as coupled parts of the same multiscale process.

This bidirectional view also changes how network findings may inform
intervention thinking. Symptom-focused and context-focused strategies need not be treated as competing alternatives. They target different parts of the same system. Symptom-focused interventions act on active symptoms and shorter-term symptom processes. Context-focused interventions act on social, economic, health-related, or lifestyle conditions that may keep symptoms easier to express over longer periods. If adverse contextual conditions persist, symptom-focused improvements may be more difficult to sustain, and the response to contextual intervention may itself depend on feedback between symptoms and context \cite{park2026precariousness}. This also implies that symptom-focused interventions may need to be adapted to the contextual conditions that help maintain vulnerability.
Conversely, improving those conditions may reduce vulnerability to future symptom expression
\cite{lund2018social,patel2018lancet,oswald2024interventions}.

This is not an argument for replacing clinical interventions with context-focused interventions. Intervention targets should match the level
at which vulnerability is being maintained. Sometimes the immediate
target will be an active symptom process. Sometimes slower contextual
conditions will be central. Often, both will matter. The slow--fast
perspective makes these possibilities easier to consider within one
framework rather than forcing a choice between them.

The present analysis remains an illustration rather than a dynamic test
of that framework. The data are cross-sectional, so they cannot
establish temporal order or causal direction and intensity of the feedback between layers.
The Ising activation parameters should likewise not be interpreted as
direct measures of contextual influence. Slow Risk Load defined the
focal comparison, making contextual conditions a theoretically relevant
source of the observed differences in baseline symptom tendency, but
those parameters may also reflect unmeasured biological, psychological,
measurement-related, or contextual factors.

The measurement and modeling choices also place limits on the
interpretation. Dichotomizing the PHQ-9 items discarded information about symptom severity, and dividing Slow Risk Load into Low and High groups simplified a continuous contextual dimension. Slow Risk Load also combines several distinct domains and therefore does not identify which contextual conditions matter most for which symptoms. Finally, the precise interaction and activation estimates depend on model specification because the two sets of parameters are statistically interdependent. For example, the sleep activation difference was negative when interactions were estimated separately across groups ($\Delta h=-0.178$), but became slightly positive when interactions were constrained to be equal ($\Delta h=0.075$), as the activation parameters readjusted under the common-interaction specification.

%

These limitations point directly to the next methodological step.
Computational models can make the slow--fast perspective explicit by
representing fast symptom dynamics, slower contextual changes, shifts in
symptom activation, and feedback between the two layers within the same
system \cite{park2026slowfast}. Such models can be used to examine how
contextual perturbations alter symptom dynamics, when symptom-to-context
feedback slows recovery, and how unmeasured contextual variation can affect
estimated symptom networks.

The present empirical analysis provides one motivation for this modeling
direction. Applied to longitudinal or intensive time-series data, slow--fast
models could be used to test whether contextual changes precede changes in
symptom expression, symptom interactions, or both, and to compare the
consequences of symptom-focused, context-focused, and combined
interventions \cite{park2026causalnet,park2026slowfast}.
Symptom networks are estimated from people living within co-evolving social, economic, psychosocial, health, and lifestyle conditions. A single-layer network is therefore not a complete description of the symptom system on its own. It is just one view
of that system under the seemingly fixed conditions in which the symptoms were observed.

\section{Methods}\label{sec:methods}

\subsection{Sample and measures}
\begin{figure}[!htbp]
\centering
\includegraphics[width=0.9\textwidth]{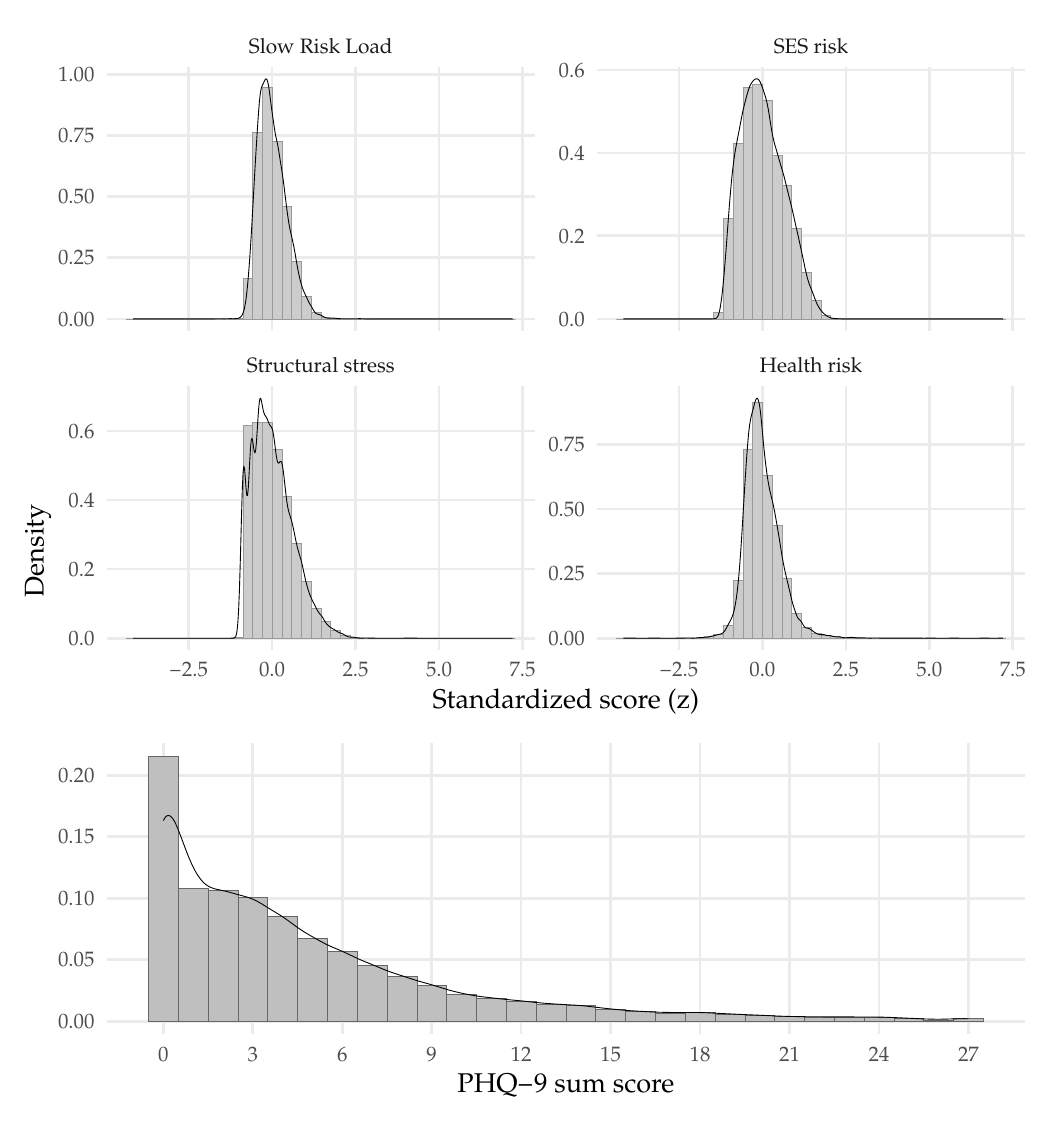}
\caption{\textbf{Distributions of slow-layer indices and depressive symptom level.}
Top panels show the empirical distributions of \textit{Slow Risk Load} and its three domain scores (SES risk, structural stress, and health risk), each expressed in standardized ($z$) units. Grey histograms are scaled to density (area = 1) and overlaid with a smoothed density curve to highlight distributional shape.
The bottom panel shows the distribution of PHQ-9 sum scores (0--27) as a density-scaled histogram (binwidth = 1, so each bar corresponds to one integer score) with an overlaid smoothed density curve. Although the PHQ-9 total is discrete, plotting density facilitates comparison of distributional shape across panels.
}
    \label{fig:slowlayer_phq_distributions}
\end{figure}

We analyzed baseline data from the HELIUS cohort, a population-based study in Amsterdam designed to investigate social and biological determinants of health \cite{snijder2017cohort}. Depressive symptoms were measured with the nine PHQ-9 symptom domains: anhedonia (\texttt{anh}), depressed mood (\texttt{dep}), sleep problems (\texttt{slp}), low energy (\texttt{ene}), appetite change (\texttt{app}), guilt/worthlessness (\texttt{glt}), concentration problems (\texttt{con}), psychomotor change (\texttt{mot}), and suicidality (\texttt{sui}).\footnote{Previous work in a multi-ethnic Dutch population found evidence for measurement invariance of the PHQ-9 with respect to ethnicity, supporting its use for comparing depressive symptom levels across ethnic groups in this context \cite{galenkamp2017measurement}.} The PHQ-9 sum score (\texttt{PHQsum}) was used as a summary measure of depressive symptom level. Table~\ref{tab:sample_summary} reports sample characteristics, Slow Risk Load and its domain scores, and symptom presence rates; Fig.~\ref{fig:slowlayer_phq_distributions} visualizes the distributions of the slow-layer indices and PHQ-9 scores in the analytic sample.

For the Ising analyses, PHQ-9 items were dichotomized to indicate symptom presence versus absence (item score $\ge 1$ vs.\ $0$), yielding binary variables $X_i \in \{0,1\}$. This dichotomization aligns the symptom measurements with the binary Ising parameterization used in the Network Comparison Test and the exact multigroup Ising models described below.

We defined three two-level groupings used throughout the Results. Ethnicity was coded as Dutch-origin versus non-Dutch-origin based on the HELIUS ethnicity variable. Age Group was defined by a median split of age (Younger vs Older). Slow Risk Group was defined by a median split of the Slow Risk Load score described below (Low vs High).

\begin{table}[t]
\caption{Characteristics of the analytic sample and variable distributions.}
\label{tab:sample_summary}
\centering
\small
\setlength{\tabcolsep}{6pt}
\renewcommand{\arraystretch}{1.08}
\begin{tabularx}{\linewidth}{@{}L r r@{}}
\toprule
\textbf{Variable} & \textbf{Summary} & \textbf{Missing} \\
\midrule
\multicolumn{3}{@{}l}{\textit{Sample characteristics}}\\
\addlinespace[2pt]
N (analytic sample)                    & 23689          & --- \\
Age (years), mean (SD)                 & 43.8 (13.4)    & 0.0\% \\
Female, n (\%)                         & 13609 (57.4\%) & 0.0\% \\
Non-Dutch, n (\%)                      & 19054 (80.4\%) & 0.0\% \\
\addlinespace[4pt]
\multicolumn{3}{@{}l}{\textit{Slow-layer composite and domain scores}}\\
\addlinespace[2pt]
Slow Risk Load, mean (SD)              & 0.01 (0.43)    & 0.0\% \\
\hspace*{1em}SES risk domain, mean (SD)            & 0.01 (0.64)    & 0.0\% \\
\hspace*{1em}Structural stress domain, mean (SD)   & 0.01 (0.62)    & 0.0\% \\
\hspace*{1em}Health risk domain, mean (SD)         & 0.00 (0.53)   & 0.0\% \\
\addlinespace[4pt]
\multicolumn{3}{@{}l}{\textit{Depression severity}}\\
\addlinespace[2pt]
PHQ-9 sum score (0--27), mean (SD)      & 4.73 (5.21)    & 0.0\% \\
\addlinespace[4pt]
\multicolumn{3}{@{}l}{\textit{PHQ-9 item presence ($\ge 1$), n (\%)}}\\
\addlinespace[2pt]
\hspace*{1em}Anhedonia (\texttt{anh})              & 12074 (51.0\%) & 0.1\% \\
\hspace*{1em}Depressed mood (\texttt{dep})         & 8208 (34.7\%)  & 0.2\% \\
\hspace*{1em}Sleep problems (\texttt{slp})         & 12013 (50.8\%) & 0.3\% \\
\hspace*{1em}Low energy (\texttt{ene})   & 15038 (63.6\%) & 0.3\% \\
\hspace*{1em}Appetite change (\texttt{app})        & 9030 (38.3\%)  & 0.5\% \\
\hspace*{1em}Guilt (\texttt{glt})                  & 5650 (23.9\%)  & 0.1\% \\
\hspace*{1em}Concentration problems (\texttt{con}) & 6779 (28.6\%)  & 0.1\% \\
\hspace*{1em}Psychomotor change (\texttt{mot})     & 4606 (19.4\%)  & 0.0\% \\
\hspace*{1em}Suicidality (\texttt{sui})            & 1838 (7.8\%)   & 0.2\% \\
\bottomrule
\end{tabularx}
\vspace{2pt}
\footnotesize\emph{Note.} "Missing" reports the percentage of missing values in the analytic sample. PHQ-9 item presence is defined as item score $\ge 1$ versus 0, matching the dichotomization used in the Ising analyses.
\end{table}

\subsection{Slow-layer construct: Slow Risk Load}

To operationalize slow-layer conditions, we constructed a composite Slow Risk Load score from indicators spanning three prespecified domains: (i) socioeconomic resources (income difficulty, work situation, labor participation, education, occupational level, health literacy, and social support); (ii) psychosocial and structural stressors (discrimination, Dutch language difficulty, cultural distance, work stress, and home stress); and (iii) health and lifestyle risks (physical health functioning, physical activity, smoking exposure, and alcohol consumption). Variables reflecting resources or protection were reverse-coded so that higher values consistently indicate more adverse slow-layer conditions. For Dutch participants, Dutch language difficulty and cultural distance were not applicable. Missing values on these indicators were therefore coded as zero on the original scale, representing no language difficulty or cultural distance, before all slow-layer indicators were standardized.

Indicators were standardized before aggregation. Domain scores were
computed as the row-wise mean of the standardized indicators within
each domain, and Slow Risk Load was computed as the mean of the three
domain scores. Because these scores are averages of standardized
indicators, they are centered near zero but do not necessarily have
unit variance. The Low and High Slow Risk groups were defined by a
median split of this composite score. A separately standardized version
of Slow Risk Load was used for plotting and for the continuous
regression analyses reported per one-standard-deviation increase.

\subsection{Network Comparison Test}

We first compared the estimated symptom networks of the Low and High
Slow Risk groups using the Network Comparison Test (NCT), a
permutation-based procedure for comparing psychological networks
\cite{van2023comparing}. The analysis was conducted
on participants with complete data for all nine binary symptom
indicators, yielding $n=11{,}717$ in the Low Slow Risk group and
$n=11{,}566$ in the High Slow Risk group.

We estimated weighted binary symptom networks separately in the two
groups using the Ising-model procedure implemented within the
\texttt{NetworkComparisonTest} package. The regularization parameter
was set to $\gamma=0.25$, and edges were retained using the AND rule.
The comparison used $10{,}000$ permutations.
The activation or threshold parameters were estimated separately
within each group. The network-structure comparison therefore did not
impose equality of symptom prevalences or activation parameters across
the Low and High Slow Risk groups.

We tested two overall network-level statistics. The network-structure statistic
$M$ is the largest absolute difference between corresponding edge
estimates in the two groups. The global-strength statistic $S$ is the
absolute difference between the sums of the absolute edge weights in
the two estimated networks. We also tested differences in individual edges. Because 36 edges were tested, the resulting \(p\)-values were adjusted for multiple comparisons using the Holm procedure. The edge-specific results are reported in Appendix Table~A4.

\subsection{Exact multigroup Ising models}

To test interaction and activation invariance separately, we fitted
exact multigroup Ising models to the same complete-case Low and High
Slow Risk samples. For group $g$, the probability of symptom
configuration $\mathbf{x}$ was specified as

\begin{equation}
P_g(\mathbf{X}=\mathbf{x})
=
\frac{
\exp\left[
\sum_{i=1}^{p}h_{ig}x_i
+
\sum_{i<j}J_{ijg}x_ix_j
\right]
}{
Z_g
},
\end{equation}

where $p=9$, $h_{ig}$ is the activation or external-field parameter
for symptom $i$ in group $g$, $J_{ijg}$ is the pairwise interaction
parameter for symptoms $i$ and $j$, and $Z_g$ is the group-specific
normalizing constant. The model used the $\{0,1\}$ parameterization
and did not include a separate inverse-temperature parameter.

The activation parameter $h_{ig}$ represents the conditional tendency
for symptom $i$ to be present when the remaining symptoms are absent.
More positive values indicate greater activation under this
parameterization. The interaction parameter $J_{ijg}$ represents the
conditional statistical association between symptoms $i$ and $j$.
Neither parameter should by itself be interpreted as a directed causal
effect.

We fitted four models that differed in the equality constraints imposed
across the two groups:

\begin{enumerate}
    \item a fully group-specific model, in which both
    $\mathbf{J}_{\mathrm{Low}}\neq\mathbf{J}_{\mathrm{High}}$ and
    $\mathbf{h}_{\mathrm{Low}}\neq\mathbf{h}_{\mathrm{High}}$;

    \item a common-interaction model, in which
    $\mathbf{J}_{\mathrm{Low}}=\mathbf{J}_{\mathrm{High}}$ but the
    activation parameters were group-specific;

    \item a common-activation model, in which the interaction
    parameters were group-specific but
    $\mathbf{h}_{\mathrm{Low}}=\mathbf{h}_{\mathrm{High}}$; and

    \item a fully invariant model, in which both interaction and
    activation parameters were constrained to equality.
\end{enumerate}

With nine symptoms, there are nine activation parameters and $\binom{9}{2}=36$ unique pairwise interaction parameters in each
group. The fully group-specific model therefore estimates both sets
separately in the two groups, giving
$2(9+36)=90$ free parameters. If interactions are common across groups, the model contains one set of 36 interaction parameters and two sets of nine activation parameters, giving
$36+2(9)=54$. If activation is common instead, the model contains two
sets of 36 interaction parameters and one set of nine activation parameters, giving $2(36)+9=81$. When both components are common, the
model contains $36+9=45$ free parameters.

With nine binary symptoms, there are only $2^9=512$ possible symptom
patterns. For each set of parameter values, the model assigns a weight
to every pattern. The partition function is the sum of these weights
across all 512 patterns and is used to normalize them into probabilities
that sum to one. We calculated this quantity exactly by enumerating the
complete state space rather than using an approximation.

Parameters were estimated by maximum likelihood using the BFGS
algorithm implemented in \texttt{optim} in R. Analytically derived
gradients of the log-likelihood were supplied to the optimizer to
improve computational efficiency and numerical accuracy.

We tested equality of the 36 interaction parameters by comparing the fully group-specific model with the common-interaction model using a
likelihood-ratio test. Equality of the nine activation parameters was tested by comparing the fully group-specific model with the
common-activation model. We additionally compared all four models using AIC and BIC. Because
the sample is large, the likelihood-ratio tests may detect statistically significant differences even when the individual parameter differences
are small. We therefore considered the likelihood-ratio tests together with the information criteria and the sizes of the estimated between-group differences.
\subsection{Activation differences and uncertainty}

To examine activation differences without imposing interaction
invariance, we used estimates from the fully group-specific model. For
each symptom $i$, the group difference was defined as

\begin{equation}
\Delta h_i
=
h_{i,\mathrm{High}}
-
h_{i,\mathrm{Low}},
\end{equation}

so that positive values indicate higher activation in the High Slow Risk
group.

For each symptom, the standard error of $\Delta h_i$ was derived from the group-specific covariance matrices of the fully group-specific model. Because the Low and High Slow Risk groups consist of independent samples, the between-group covariance is zero, so
\[
\mathrm{Var}(\Delta h_i)
=
\mathrm{Var}(h_{i,\mathrm{High}})
+
\mathrm{Var}(h_{i,\mathrm{Low}}).
\]
We then constructed 95\% Wald confidence intervals as $\Delta h_i \pm 1.96\,\mathrm{SE}(\Delta h_i)$.

For symptom-specific inference, we additionally calculated two-sided Wald $p$-values for the nine activation contrasts and adjusted these for multiple comparisons using the Holm procedure. The unadjusted confidence intervals are retained to show the magnitude and uncertainty of the estimated differences,
whereas statistical evidence for individual symptom differences is evaluated using the Holm-adjusted $p$-values.

As a sensitivity analysis, we also calculated activation differences
under the common-interaction model, in which the interaction parameters
were constrained to equality while activation parameters remained
group-specific. This comparison assessed whether the overall activation
pattern depended on imposing a common interaction matrix. The complete
activation estimates under both specifications are reported in Appendix
Tables~A2 and~A3.

\subsection{Symptom-specific associations with Slow Risk Load}
To examine whether Slow Risk Load was associated uniformly with all depressive symptoms or more strongly with particular symptoms, we estimated symptom-wise logistic regression models for each PHQ-9 item. Outcomes were binary symptom indicators constructed from the original 0--3 response scale, with $1$ indicating any symptom presence ($>0$) and $0$ indicating absence.

For each symptom $i$, we first estimated a covariate-adjusted model:
\[
\text{logit}\,\Pr(X_i=1)
=
\beta_{0i}
+
\beta_{1i}\,\text{SlowRiskLoad}
+
\beta_{2i}\,\text{Age}
+
\beta_{3i}\,\text{Gender}
+
\beta_{4i}\,\text{Ethnicity}.
\]
This model estimates the association between Slow Risk Load and symptom $i$ while adjusting for age, gender, and ethnicity.

We then estimated a symptom-adjusted model for each symptom:
\[
\text{logit}\,\Pr(X_i=1)
=
\beta_{0i}
+
\beta_{1i}\,\text{SlowRiskLoad}
+
\beta_{2i}\,\text{Age}
+
\beta_{3i}\,\text{Gender}
+
\beta_{4i}\,\text{Ethnicity}
+
\sum_{j\neq i}\theta_{ij}X_j .
\]
This model additionally adjusts for the other eight depressive symptoms. The coefficient $\beta_{1i}$ therefore reflects the association between Slow Risk Load and symptom $i$ after accounting for the remaining symptoms.

Comparing the covariate-adjusted and symptom-adjusted estimates provides a descriptive view of how the association between Slow Risk Load and each symptom changes after conditioning on the remaining symptom profile. Because those symptoms may themselves be downstream of Slow Risk Load or of one another, the symptom-adjusted coefficients are not interpreted as direct causal effects.

Slow Risk Load was standardized before analysis, so $\beta_{1i}$ is the log-odds change in symptom presence per 1 SD increase in Slow Risk Load. For Fig.~\ref{fig:slowrisk_coefficients}, we report the $\beta_{1i}$ estimates from both models with 95\% confidence intervals.

\section*{Data and code availability}

Analysis code and the non-sensitive derived outputs used to generate
the reported tables and figures are available at
\url{https://github.com/KyuriP/slow-fast-perspective}. Individual-level
HELIUS data cannot be shared publicly because of participant privacy
and study-governance requirements. Researchers may apply for access
through the HELIUS study's established data-access procedures.


\FloatBarrier
\clearpage
\appendix

\setcounter{table}{0}
\setcounter{figure}{0}

\renewcommand{\thetable}{A\arabic{table}}
\renewcommand{\thefigure}{A\arabic{figure}}

\floatplacement{table}{!htbp}

\section{Additional results and numerical diagnostics}
\label{app:additional_results}

\subsection{Optimization diagnostics for the exact Ising models}

The partition function for each exact multigroup Ising model was
evaluated over all $2^9=512$ possible symptom configurations.
Parameters were estimated by maximum likelihood using the BFGS
algorithm with analytic gradients.

Table~A1 reports numerical diagnostics
for the four fitted models. All models terminated with convergence
code zero. Their Hessian matrices were positive definite and
invertible, allowing the corresponding covariance matrices to be used
for Wald standard errors and confidence intervals.

\begin{table}
\caption{\label{tab:tab:appendix_optimization}Optimization diagnostics for the four exact multigroup Ising models. A convergence code of zero indicates successful termination.}
\centering
\begin{tabular}[!htbp]{lcccccc}
\toprule
Model & $k$ & Code & Max. $|g|$ & \shortstack{Min. Hessian\\eigenvalue} & \shortstack{Condition\\number} & Covariance\\
\midrule
Fully group-specific & 90 & 0 & 8.51e-04 & 21.53 & 2367.72 & Yes\\
Common interactions & 54 & 0 & 3.25e-05 & 37.81 & 1726.48 & Yes\\
Common activation & 81 & 0 & 4.36e-05 & 26.94 & 2001.15 & Yes\\
Fully invariant & 45 & 0 & 5.01e-04 & 58.48 & 1401.76 & Yes\\
\bottomrule
\end{tabular}
\end{table}

\subsection{Activation estimates from the fully group-specific model}

Table~A2 reports the activation
parameters from the fully group-specific exact Ising model. Both
interaction and activation parameters were estimated separately in
the Low and High Slow Risk groups. The contrast was defined as

\[
\Delta h_i
=
h_{i,\mathrm{High}}
-
h_{i,\mathrm{Low}},
\]

so that positive values indicate higher activation in the High Slow
Risk group when the remaining symptoms are absent.

\begin{table}[htbp]

\caption{\label{tab:tab:appendix_activation_free}
Symptom-activation estimates from the fully group-specific exact Ising model.
Both interaction and activation parameters were estimated separately in the
Low and High Slow Risk groups.}

\centering
\small
\begin{tabular}{lrrrrrrr}
\toprule
Symptom & $h_{\mathrm{Low}}$ & $h_{\mathrm{High}}$ & $\Delta h$ & SE &
95\% CI & $p$ & $p_{\mathrm{Holm}}$ \\
\midrule
Anhedonia
& -2.188 & -1.891 & 0.296 & 0.067 & [0.166, 0.427] & $<.001$ & $<.001$ \\

Depressed mood
& -3.524 & -3.158 & 0.366 & 0.097 & [0.175, 0.556] & $<.001$ & .001 \\

Sleep problems
& -1.738 & -1.916 & -0.178 & 0.063 & [-0.301, -0.055] & .005 & .028 \\

Low energy
& -1.278 & -1.204 & 0.074 & 0.053 & [-0.029, 0.177] & .163 & .196 \\

Appetite change
& -2.584 & -2.429 & 0.155 & 0.077 & [0.005, 0.306] & .044 & .176 \\

Guilt
& -3.403 & -3.215 & 0.188 & 0.102 & [-0.012, 0.387] & .065 & .196 \\

Concentration problems
& -3.029 & -2.868 & 0.161 & 0.090 & [-0.015, 0.337] & .074 & .196 \\

Psychomotor change
& -4.021 & -3.460 & 0.561 & 0.121 & [0.323, 0.799] & $<.001$ & $<.001$ \\

Suicidality
& -5.332 & -4.858 & 0.473 & 0.212 & [0.058, 0.889] & .026 & .128 \\
\bottomrule
\end{tabular}

\vspace{0.5em}
\begin{minipage}{0.98\linewidth}
\footnotesize
\textit{Note.} $\Delta h = h_{\mathrm{High}} - h_{\mathrm{Low}}$, such that
positive values indicate higher activation in the High Slow Risk group.
Two-sided Wald $p$-values are based on the symptom-specific activation contrasts.
$p_{\mathrm{Holm}}$ denotes adjustment across the nine contrasts using the Holm
procedure. Confidence intervals are unadjusted 95\% Wald confidence intervals.
\end{minipage}

\end{table}

\subsection{Sensitivity to the interaction specification}

The main analysis allowed both interaction and activation parameters
to differ across the Low and High Slow Risk groups. As a sensitivity
analysis, we compared those activation differences with estimates
from the model in which the 36 interaction parameters were constrained
to equality while the nine activation parameters remained
group-specific.

\begin{figure}[htbp]
\centering
\includegraphics[width=1\textwidth]
{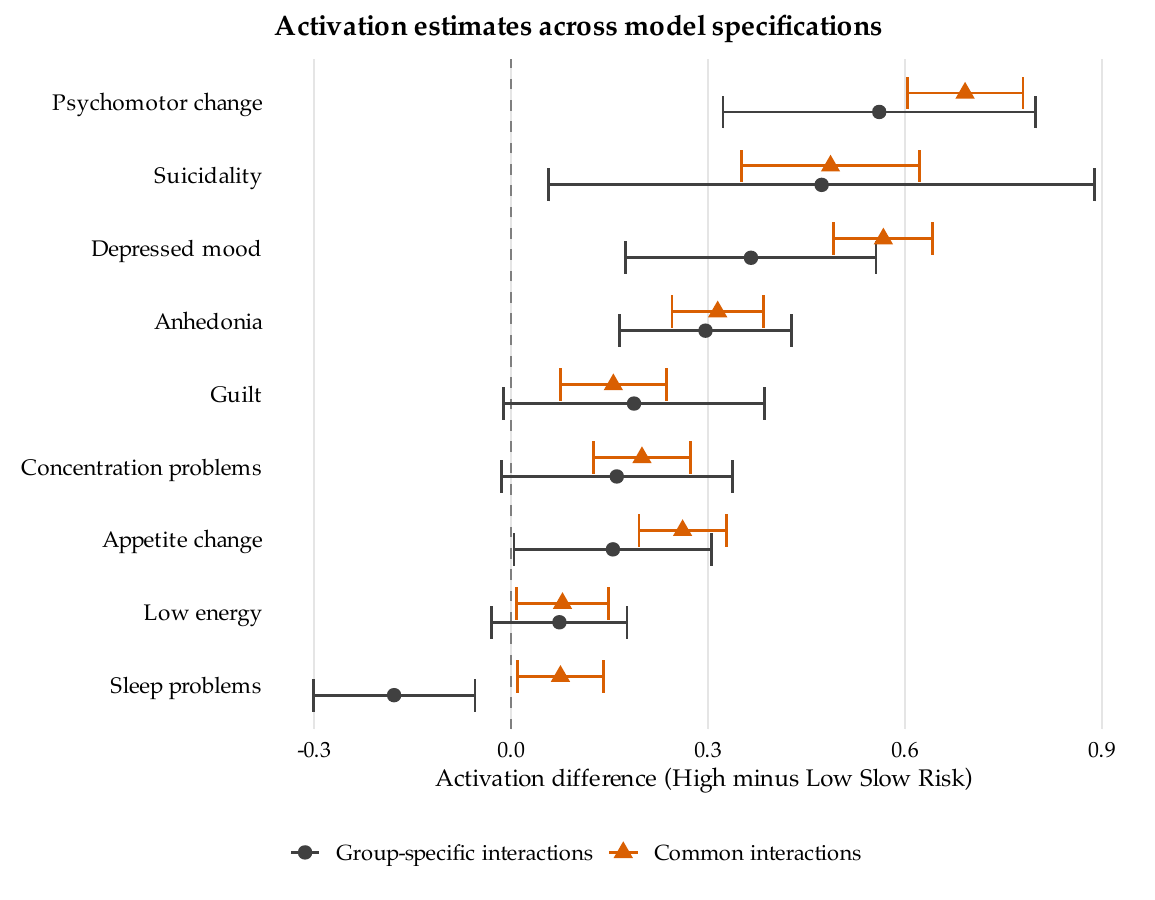}
\caption{\textbf{Sensitivity of symptom-activation differences to the
interaction specification.}
Points show activation-parameter differences,
$\Delta h_i=h_{i,\mathrm{High}}-h_{i,\mathrm{Low}}$, from the fully
group-specific model and the common-interaction model. Horizontal
bars show 95\% Wald confidence intervals. Positive values indicate
higher activation in the High Slow Risk group when the remaining
symptoms are absent.}
\label{fig:appendix_activation_sensitivity}
\end{figure}

Figure~\ref{fig:appendix_activation_sensitivity} shows that the broad
tendency toward higher activation in the High Slow Risk group was
present under both specifications. The symptom-specific estimates
nevertheless changed, most visibly for sleep problems. This comparison
supports the overall activation contrast while also illustrating the
statistical interdependence of interaction and activation estimates.

Table~A3 reports the corresponding
activation estimates from the common-interaction model.

\begin{table}

\caption{\label{tab:tab:appendix_activation_commonJ}Symptom-activation estimates from the exact Ising model with interactions constrained to equality and activation parameters estimated separately across groups.}
\centering
\begin{tabular}[H]{lllllll}
\toprule
Symptom & $h_{\mathrm{Low}}$ & $h_{\mathrm{High}}$ & $\Delta h$ & SE & 95\% CI lower & 95\% CI upper\\
\midrule
Anhedonia & -2.197 & -1.882 & 0.315 & 0.035 & 0.245 & 0.384\\
Depressed mood & -3.628 & -3.060 & 0.567 & 0.038 & 0.492 & 0.643\\
Sleep problems & -1.838 & -1.762 & 0.075 & 0.033 & 0.010 & 0.141\\
Low energy & -1.277 & -1.198 & 0.079 & 0.036 & 0.008 & 0.149\\
Appetite change & -2.639 & -2.378 & 0.262 & 0.034 & 0.195 & 0.328\\
Guilt & -3.398 & -3.242 & 0.156 & 0.041 & 0.075 & 0.237\\
Concentration problems & -3.053 & -2.854 & 0.200 & 0.038 & 0.126 & 0.274\\
Psychomotor change & -4.106 & -3.415 & 0.692 & 0.045 & 0.604 & 0.780\\
Suicidality & -5.361 & -4.874 & 0.487 & 0.069 & 0.351 & 0.622\\
\bottomrule
\end{tabular}
\end{table}

\subsection{Edge-specific Network Comparison Test results}

The omnibus Network Comparison Test did not reject invariance of the
overall network structure between the Low and High Slow Risk groups.
For transparency, Table~A4 reports all 36
edge-specific comparisons. These results are treated as secondary to
the omnibus structure test. Both unadjusted permutation $p$-values and
Holm-adjusted $p$-values are reported.

\begingroup
\small
\setlength{\tabcolsep}{4.5pt}
\renewcommand{\arraystretch}{1.08}

\begin{longtable}{@{}llrrrrrr@{}}
\caption{Edge-specific Network Comparison Test results for the Low and
High Slow Risk groups.}
\label{tab:appendix_nct_edges}\\

\toprule
Symptom 1
& Symptom 2
& Low
& High
& $\Delta J$
& $|\Delta J|$
& $p$
& $p_{\mathrm{Holm}}$ \\
\midrule
\endfirsthead

\multicolumn{8}{@{}l}{\textit{Table A4 continued}}\\
\toprule
Symptom 1
& Symptom 2
& Low
& High
& $\Delta J$
& $|\Delta J|$
& $p$
& $p_{\mathrm{Holm}}$ \\
\midrule
\endhead

\midrule
\multicolumn{8}{r@{}}{\textit{Continued on next page}}\\
\endfoot

\bottomrule
\endlastfoot

Guilt & Psychomotor change & 0.471 & 0.782 & 0.311 & 0.311 & $<.001$ & .036\\
Depressed mood & Guilt & 1.339 & 1.020 & -0.319 & 0.319 & .002 & .070\\
Depressed mood & Sleep problems & 0.505 & 0.732 & 0.227 & 0.227 & .006 & .204\\
Appetite change & Concentration problems & 0.311 & 0.522 & 0.211 & 0.211 & .009 & .297\\
Guilt & Concentration problems & 0.560 & 0.786 & 0.227 & 0.227 & .013 & .416\\
Sleep problems & Appetite change & 0.520 & 0.702 & 0.182 & 0.182 & .014 & .434\\
Sleep problems & Low energy & 1.266 & 1.421 & 0.155 & 0.155 & .047 & 1.000\\
Low energy & Concentration problems & 0.787 & 0.610 & -0.177 & 0.177 & .089 & 1.000\\
Sleep problems & Psychomotor change & 0.417 & 0.248 & -0.169 & 0.169 & .116 & 1.000\\
Anhedonia & Depressed mood & 1.732 & 1.883 & 0.151 & 0.151 & .120 & 1.000\\
Depressed mood & Suicidality & 1.630 & 1.333 & -0.297 & 0.297 & .129 & 1.000\\
Anhedonia & Sleep problems & 0.435 & 0.547 & 0.112 & 0.112 & .134 & 1.000\\
Depressed mood & Appetite change & 0.406 & 0.288 & -0.117 & 0.117 & .138 & 1.000\\
Sleep problems & Suicidality & 0.000 & 0.277 & 0.277 & 0.277 & .193 & 1.000\\
Anhedonia & Appetite change & 0.655 & 0.556 & -0.099 & 0.099 & .209 & 1.000\\
Low energy & Guilt & 0.418 & 0.278 & -0.140 & 0.140 & .232 & 1.000\\
Appetite change & Guilt & 0.453 & 0.553 & 0.099 & 0.099 & .250 & 1.000\\
Low energy & Psychomotor change & 0.461 & 0.605 & 0.144 & 0.144 & .330 & 1.000\\
Depressed mood & Low energy & 0.420 & 0.529 & 0.109 & 0.109 & .331 & 1.000\\
Psychomotor change & Suicidality & 0.773 & 0.888 & 0.115 & 0.115 & .379 & 1.000\\
Anhedonia & Psychomotor change & 0.212 & 0.116 & -0.096 & 0.096 & .403 & 1.000\\
Appetite change & Psychomotor change & 0.391 & 0.465 & 0.074 & 0.074 & .426 & 1.000\\
Depressed mood & Psychomotor change & 0.581 & 0.655 & 0.073 & 0.073 & .451 & 1.000\\
Sleep problems & Guilt & 0.286 & 0.355 & 0.069 & 0.069 & .499 & 1.000\\
Anhedonia & Concentration problems & 0.420 & 0.450 & 0.030 & 0.030 & .764 & 1.000\\
Low energy & Appetite change & 1.117 & 1.142 & 0.025 & 0.025 & .782 & 1.000\\
Sleep problems & Concentration problems & 0.452 & 0.475 & 0.023 & 0.023 & .791 & 1.000\\
Depressed mood & Concentration problems & 0.434 & 0.415 & -0.019 & 0.019 & .807 & 1.000\\
Concentration problems & Psychomotor change & 1.364 & 1.350 & -0.014 & 0.014 & .866 & 1.000\\
Appetite change & Suicidality & 0.248 & 0.226 & -0.022 & 0.022 & .872 & 1.000\\
Anhedonia & Guilt & 0.408 & 0.397 & -0.011 & 0.011 & .921 & 1.000\\
Concentration problems & Suicidality & 0.444 & 0.434 & -0.010 & 0.010 & .930 & 1.000\\
Anhedonia & Low energy & 1.289 & 1.286 & -0.003 & 0.003 & .963 & 1.000\\
Guilt & Suicidality & 1.409 & 1.407 & -0.002 & 0.002 & .995 & 1.000\\
Anhedonia & Suicidality & 0.000 & 0.000 & 0.000 & 0.000 & 1.000 & 1.000\\
Low energy & Suicidality & 0.000 & 0.000 & 0.000 & 0.000 & 1.000 & 1.000\\

\end{longtable}

\vspace{-1em}
\begin{minipage}{\textwidth}
\footnotesize
\textit{Note.} The signed difference is High minus Low Slow Risk.
Both unadjusted and Holm-adjusted permutation \(p\)-values are shown.
Edge-specific tests are secondary to the omnibus network-structure test.
\end{minipage}
\endgroup

\clearpage

\backmatter

\bibliography{sn-bibliography}

\end{document}